\documentclass[12pt]{article}
\pdfoutput=1
\usepackage{latexsym, graphicx} 
\usepackage{amsmath}
\usepackage{amssymb}

\usepackage[colorlinks=true, linkcolor=blue, bookmarks=true]{hyperref}

\makeatletter
\renewcommand\section{\@startsection {section}{1}{\z@}%
                                   {-3.5ex \@plus -1ex \@minus -.2ex}
                                   {2.3ex \@plus.2ex}%
                                   {\normalfont\large\bfseries}}
\renewcommand\subsection{\@startsection{subsection}{2}{\z@}%
                                     {-3.25ex\@plus -1ex \@minus -.2ex}%
                                     {1.5ex \@plus .2ex}%
                                     {\normalfont\bfseries}}
\makeatother
\newcommand{\jr}{}
\newcommand{\othercit}{}

\begin{document}
\begin{center}
{\LARGE\bf AdS (in)stability: an analytic approach}    \\
\vskip 5mm
{\large Ben Craps$^a$ and Oleg Evnin$^{b,a}$}
\vskip 5mm
{\em $^a$ Theoretische Natuurkunde, Vrije Universiteit Brussel and\\
The International Solvay Institutes\\ Pleinlaan 2, B-1050 Brussels, Belgium}
\vskip 3mm
{\em $^b$ Department of Physics, Faculty of Science, Chulalongkorn University,\\
Thanon Phayathai, Pathumwan, Bangkok 10330, Thailand}

\vskip 3mm
{\small\noindent  {\tt Ben.Craps@vub.ac.be, oleg.evnin@gmail.com	}}
\vskip 5mm
\end{center}

\noindent\textbf{Abstract:} We briefly review the topic of AdS (in)stability, mainly focusing on a recently introduced analytic approach and its interplay with numerical results.\vspace{1cm}

Stability considerations are a crucial part of any comprehensive dynamical analysis. While some instabilities are catastrophic in the sense of completely destroying the unstable dynamical regime and rendering the corresponding solutions unphysical, other instabilities may only produce localized strong deviations from the original background while having little effect outside that localized region. It is instabilities of the latter sort that will concern us here.

Maximally symmetric space-times (Minkowski, de Sitter, anti-de Sitter) have a special significance in geometrical theories of gravity, providing ``empty'' arenas on which other processes can unfold. The basic question of stability for these simplest possible space-times turns out to produce some surprisingly elaborate consequences.

Before turning to our main question whether the anti-de Sitter space-time (AdS) is stable, and in particular to the analytic approach introduced in \cite{Craps:2014vaa, Craps:2014jwa, Craps:2015iia, Craps:2015xya}, let us briefly discuss the fate of Minkowski space with small perturbations excited. Consider, for instance, an infalling spherical shell of a massless scalar field minimally coupled to gravity. For sufficiently large amplitude, this leads to black hole formation. For sufficiently small amplitude, however, the shell scatters outward and disperses to infinity. Gravitational interactions effectively turn off as the shell expands to infinity, and the story basically ends here. A small amplitude shell cannot possibly produce any noticeable lasting effect. More generally, Minkowski space has been rigorously shown to be stable under small perturbations \cite{Christodoulou:1993uv}.

AdS space-time (which is a sort of space-time hyperboloid with a constant negative curvature) is known to act as a box. Massless field waves propogate out to infinity in finite time, get reflected by the (conformal) boundary and return to the interior. This crucial feature eliminates the dispersion mechanism described above underlying the stability of Minkowski space. If a shell has too small amplitude to form a black hole right away, it scatters outward to the boundary. But the boundary will reflect the shell, which can keep trying to form a black hole, slightly changing its shape every time. It is a nontrivial question whether this eventually results in black hole formation.

While considering the evolution of the AdS perturbations of the sort we have described is interesting in itself from a mathematical general relativity point of view, the notion of gravitational holography provides additional motivation. The AdS/CFT correspondence connects the dynamics of AdS perturbations to the dynamics of a certain quantum conformal field theory living in one spatial dimension less. A black hole in the interior of AdS often corresponds in this context to a thermal state in a dual conformal field theory, so the process of black hole formation is dual to thermalization.

In 2011, Bizo\'n and Rostworowski \cite{Bizon:2011gg} studied Gaussian shells with amplitude $\epsilon$ (and some specific width), and numerically found black hole formation at times scaling like $1/\epsilon^2$. The mechanism behind the black hole formation is the turbulent flow of energy to short wavelengths, which is eventually cut off by the formation of a horizon. Naive perturbation theory in $\epsilon$ indeed shows energy transfer to high frequencies, but really breaks down at times of order $1/\epsilon^2$. On the other hand, for special initial data, with only one normal mode excited, both perturbation theory and numerics indicate regular evolution, at least up to times of order $1/\epsilon^2$.

A highly special object from a mathematical perspective, the AdS space-time possesses a number of peculiar properties that contribute to the dynamics of its perturbations. All classical particles emitted from a point refocus to the same point after a certain period (related to the AdS curvature). Similarly, solutions to the wave equations display simple periodicity properties. Once a shell-like perturbation expands enough so that the interactions are effectively turned off, its propagation in the underlying AdS background will simply refocus it to the original shape. After that, the gravitational interactions will have one more chance to make the shell more compact. (These reflection properties of the AdS boundary are very special. In generic situation, there is no reason to expect that wave profile focusing will not be upset by propagating out and getting reflected back.) Further discussion of shell propagation from such viewpoint can be found in \cite{Dimitrakopoulos:2014ada}.

Numerical evidence suggests that certain perturbations of AdS lead to instability through black hole formation, no matter how small the perturbation amplitude is. In order for a small perturbation to form a black hole, its energy must become focused in a tiny region. This is only possible if there is an efficient way to transfer excitation energy to arbitrarily short-wavelength modes. Such energy transfer is generally known as turbulence, and when it is driven only by non-linearities (rather than by linear instabilities), it is known as weak turbulence. One thus has to deal in this context with the interrelated notions of weak turbulence, non-linear instability and gravitational collapse.

Let us look more closely into the perturbative study of a scalar field in AdS$_{d+1}$ \cite{Bizon:2011gg}. Consider a spherically symmetric scalar field $\phi(x,t)$, minimally coupled to the metric
\begin{equation}
ds^{2}=\frac{L^{2}}{\cos^{2}x}\left(\frac{dx^{2}}{ A(x,t)}-A(x,t)e^{-2\delta(x,t)}dt^{2}+\sin^{2}x\,d\Omega_{d-1}^{2}\right).
\label{metric}
\end{equation}
Unperturbed global AdS, with a boundary at $x=\pi/2$, corresponds to $A=1$ and $\delta=0$. Because of spherical symmetry, the metric is fixed by the constraints on any given time slice, so all we need to do is solve the equation of motion for $\phi$. We introduce a perturbative expansion 
\begin{equation}\label{pert}
\phi=\epsilon\phi_{(1)}+\epsilon^3\phi_{(3)}+\ldots
\end{equation}
(It is straightforward to check that, for a minimally coupled scalar field, the metric is only corrected at even orders in $\epsilon$ and the scalar field at odd orders.) The first term $\phi_{(1)}$ is a solution to the linearized equation of motion, i.e., that of a free scalar in AdS; it can be expanded in normal modes $e_n(x)$,
\begin{equation}\label{phi1}
\phi_{(1)}(x,t)=\sum_{n=0}^{\infty}a_{n}\cos(\omega_{n}t+b_{n})e_{n}(x),
\end{equation}
with frequencies given by
\begin{equation}\label{freqs}
\omega_n=d+2n.
\end{equation}
The frequencies are all integers and the spectrum is ``fully resonant.'' This highly special feature of AdS will play a crucial role in the weakly nonlinear dynamics of small AdS perturbations. The integer character of the spectrum would not have been upset if we had included non-sperically-symmetric modes, or used a non-scalar field (while including a non-zero mass would have shifted the spectrum by a non-integer value, leaving differences of any two frequencies integer). Note that the opposite case of non-resonant spectra ($\sum_i n_i\omega_i\ne 0$ for any integer $n_i$) 
forms the basis for the celebrated Kolmogorov-Arnol'd-Moser theory, which rigorously states that sufficiently small perturbations cannot have any significant effect on the dynamics in such a setting for all times. It is the resonant character of the AdS spectrum that makes it vulnerable to small perturbations. (The Kolmogorov-Arnol'd-Moser results technically impose some ``diophantine'' conditions of the form $\left|\sum_i n_i \omega_i\right| > \mathcal{O}(1/(\sum_i|n_i|)^\alpha))$ for all integers $n_i$ and some $\alpha$. For example, in two dimensions, this condition simply says that the ratio of the two normal mode frequencies is sufficiently irrational and not efficiently approximated by rational numbers.  Such conditions are met by typical ``algebraic numbers'' solving polynomial equations with integer coefficients, $\sqrt{2}$ or $\sqrt{2+\sqrt{5}}$ for instance, but can be violated by some transcendental numbers, like Liouville's number $\sum_n 10^{-n!}$. The situation becomes more subtle in an infinite number of dimensions where a non-resonant spectrum may approach resonant values asymptotically in the ultraviolet. Some further comments on the  Kolmogorov-Arnol'd-Moser theory and related results in the context of AdS stability can be found in \cite{Menon:2015oda}.)

To proceed further, it is convenient to expand $\phi_{(3)}$ in the same set of linearized normal modes,
\begin{equation}\label{phi3}
\phi_{(3)}(x,t)=\sum_{n=0}^{\infty}c_{n}(t)e_{n}(x).
\end{equation}
The expansion coefficients $c_n(t)$ are determined by the equations of motion
\begin{equation}
\ddot{c}_{n}+\omega_{n}^{2}c_{n}=
\Omega_{ijkn}a_{i}a_{j}a_{k}\cos[(\omega_{i}\pm\omega_{j}\pm\omega_{k})t+(b_{i}\pm b_{j}\pm b_{k})]+\ldots\ ,
\end{equation}
where $\Omega_{ijkn}$ are specific complicated integrals involving AdS mode functions \cite{Craps:2014vaa, Craps:2014jwa}. 

Driving frequencies $\omega_i\pm\omega_j\pm\omega_k$ different from $\omega_n$ lead to oscillatory contributions, which remain perturbatively small for all times and can therefore be ignored. More dangerous contributions come from resonances,
\begin{equation}\label{res}
\pm\omega_n=\omega_i\pm\omega_j\pm\omega_k, 
\end{equation}
which are numerous because of the integer normal mode spectrum (\ref{freqs}). Resonances lead to secular terms
\begin{equation}
c_{n}(t)=\  \Omega_{ijkn}a_{i}a_{j}a_{k}\,t\sin(\omega_{n}t+(b_{i}\pm b_{j}\pm b_{k}))+\ldots\ ,
\end{equation}
which invalidate naive perturbation theory on time scales of order $1/\epsilon^2$. (The first subleading correction $\phi^{(3)}$  becomes of the same order as the leading contribution $\phi^{(1)}$.)

Secular terms can be illustrated via a simpler example, namely a particle in an anharmonic oscillator potential
\begin{equation}
V(x)=\frac{\omega^{2}x^{2}}{2}+\frac{x^{4}}{4},
\end{equation}
leading to the equation of motion $\ddot x+\omega^2 x+x^3=0$. This equation can be treated perturbatively by expanding $x(t)=\epsilon x_1(t)+\epsilon^3 x_3(t)+\ldots$ and solving order by order in $\epsilon$:
\begin{equation}
x(t)=\epsilon\, a\cos(\omega t+b)+\epsilon^3\left(\frac{a^{3}}{32\omega^{2}}\cos(3\omega t+3b)-\frac{3a^{3}}{8\omega}t\sin(\omega t+b)\right)+\mathcal{O}\left(\epsilon^{5}\right).
\end{equation}
Comparing with the exact solution (obtained, for example, by numerical integration), one finds that the zeroth order solution works well at early times, but drifts completely out of phase with the exact solution on time scales of order $1/\epsilon^2$. The first order correction further improves the zeroth order solution at early times, but causes the perturbative solution to diverge away from the exact solution at times of order $1/\epsilon^2$, because of the secular term proportional to $t$. The perturbation theory can be resummed, however, by absorbing the secular term in the phase of the zeroth order oscillator,
\begin{equation}
x(t)=\epsilon\,a\cos\left(\omega t+b+\frac{3a^{2}}{8\omega^{2}}\epsilon^2 t\right)+\frac{\epsilon^3\,a^{3}}{32\omega^{2}}\cos\left(3\omega t+3\left(b+\frac{3a^{2}}{8\omega^{2}}\epsilon^2 t\right)\right)+\mathcal{O}\left(\epsilon^{5}\right),
\end{equation}
and this renders first order perturbation theory excellent on time scales of order $1/\epsilon^2$. Since the phase shift is proportional to $t$, it actually corresponds to a small frequency shift. This procedure is known as the Poincar\'e-Lindstedt method for resumming  (certain types of) secular terms. Note that the fact that the secular contribution could be absorbed in a frequency shift crucially depends on the particular phase of the secular term $t\sin(\omega t+b)$. It would not have worked, for example, for a secular term proportional to $t\cos(\omega t+b)$, which happens to be absent for this particular very simple problem.

Returning to the AdS problem, using frequency shifts is sufficient for resumming secular terms only for specially chosen, fine-tuned initial conditions. These initial conditions correspond to time-periodic solutions dominated by a single normal mode, which can be constructed to all orders in $\epsilon$ \cite{Bizon:2011gg, Dias:2011ss, Maliborski:2013jca}. Numerical general relativity indicates that these solutions are stable; they belong to islands of stability in the space of initial conditions \cite{Dias:2012tq, Maliborski:2013jca,  Buchel:2013uba, Maliborski:2014rma}.

This brings us to our first main message: {\em Numerics and perturbation theory suggest that arbitrarily small scalar perturbations (of size $\epsilon$) of AdS can form black holes in times of order $1/\epsilon^2$. Non-collapsing initial data also exist.}

Frequency shifts are often not enough, though. The secular terms appearing in the AdS instability problem take the form
\begin{align}
c_{n}(t)&=\ \Omega_{ijkn}a_{i}a_{j}a_{k}\,t\sin(\omega_{n}t+(b_{i}\pm b_{j}\pm b_{k}))+\ldots\nonumber\\
&=(\ldots)t\sin(\omega_n t+b_n)+(\ldots)t\cos(\omega_n t+b_n)+\ldots \label{cn}
\end{align}
The second term on the RHS, which is present for nontrivial resonances, cannot be absorbed in frequency shifts \cite{Bizon:2011gg}. (For trivial resonances $\omega_n=\omega_n+\omega_k-\omega_k$, the phase $b_{i}\pm b_{j}\pm b_{k}$ in the first line of (\ref{cn}) becomes $b_{n}+ b_{k}- b_{k}=b_n$, so in the second line only the first time is present. These special terms correspond to slow phase drifts similar to what we have seen in the simple anharmonic oscillator problem.)

In order to deal with more general secular terms, several more general resummation methods have been used for the AdS instability problem: multiscale analysis \cite{Balasubramanian:2014cja}, renormalization \cite{Craps:2014vaa} and time-averaging \cite{Basu:2014sia, Craps:2014jwa}. Regardless of which method is used, the result will be that the secular terms are replaced by a slow time-dependence of the integration constants in the lowest order (linearized) solution, and that this time-dependence is governed by flow equations.

Here we focus on the method of time-averaging \cite{Craps:2014jwa}. Consider the nonlinear equation $\ddot c_n+\omega_n^2 c_n=S_n(c)$, and approximate the source $S_n$ as cubic. (For spherically symmetric AdS perturbations, the metric can be integrated out using the constraints, and one obtains a nonlocal, nonlinear action for the scalar field, i.e., a system of nonlinearly coupled oscillators. The lowest nonlinearity makes a cubic contribution to $S_n$, but higher order corrections can become important if perturbations do not remain small, e.g., when a black hole forms. Some general bounds on errors due to such non-linearity truncation have been considered in \cite{Dimitrakopoulos:2015pwa}. In the context of AdS stability study, such bounds are aimed precisely at establishing the validity of approximate truncated equations in situations when their solutions fail to collapse.) In Hamiltonian form, the equation can be written as $\dot c_n=\pi_n$ and $\dot\pi_n=-\omega_n^2 c_n+S_n(c)$. In terms of ``interaction picture'' complex variables $\alpha_n(t)$, which would be constant in the absence of nonlinearities,  
\begin{align}
c_n&=\epsilon(\alpha_n e^{-i\omega_n t}+\bar\alpha_n e^{i\omega_n t}),\\
\pi_n&=-i\epsilon\omega_n(\alpha_n e^{-i\omega_n t}-\bar\alpha_n e^{i\omega_n t}),
\end{align}
this becomes $\dot\alpha_n=\epsilon^2 S_n(\alpha,\bar\alpha,t)$. The flow equations can be obtained by averaging $S_n$ over its explicit time-dependence: $\dot\alpha_n=\epsilon^2 S_n(\alpha,\bar\alpha)_{\rm av}$, which can be thought of as integrating out fast oscillations. It can be shown that the flow equations provide a reliable approximation on time intervals of order $1/\epsilon^2$ (unless amplitude growth were to invalidate the cubic approximation of $S_n$). It can also be shown that explicit resummation of the secular terms in naive perturbation theory using  multiscale analysis or renormalization would have produced a slow variation of the integration constants of the lowest order (linearized) solution exactly identical at this order in $\epsilon$ to the one described by the time-averaged equations introduced above.

To summarize, we have resummed the naive perturbation series (\ref{pert}) with (\ref{phi1}, \ref{phi3}, \ref{cn}) into 
\begin{equation}
\phi(x,t)=\sum_{n=0}^{\infty}\left[\epsilon\, a_{n}(\epsilon^2 t)\cos(\omega_{n}t+b_{n}(\epsilon^2 t))+\ldots\right]e_{n}(x),
\end{equation}
where the slow time-dependence of the integration constants is governed by flow equations of the form $\dot a_n=\epsilon^2(\ldots),\ \dot b_n=\epsilon^2(\ldots)$. (The relation with the complex amplitudes is $\alpha_n=a_ne^{-ib_n}/2$.) The flow equations are reliable on time intervals of order $1/\epsilon^2$, unless they cause the amplitudes to become so large that higher nonlinearities can no longer be ignored, making our cubic approximation for the source fail.

In the resonance relation (\ref{res}) all signs are independent. Given the integer spectrum (\ref{freqs}), one may have expected various combinations of signs to lead to resonant transfer of energy. Explicit computation, however, shows that all combinations of signs except $\omega_n=\omega_i+\omega_j-\omega_k$ are dynamically forbidden \cite{Craps:2014vaa}. (Curiously, the surviving combinations are those that would also be resonant for a massive scalar field.) 

The explicit form of the flow equations is 
\begin{equation}
\frac{d\alpha_n}{d\tau}=\frac{i\epsilon^2}{\omega_n}\frac{\partial W}{\partial\bar\alpha_n},
\end{equation}
with
\newcommand{\ssty}{\scriptstyle}
\begin{align}
W=\frac{1}{4}\hspace{-5mm}\sum\limits_{\begin{array}{c}\vspace{-7mm}\\\ssty ijkl\vspace{-2mm}\\\ssty\omega_{i}+\omega_{j}=\omega_{k}+\omega_{l}\end{array}}\hspace{-8mm}
\Omega_{ijkl}\alpha_{i}\alpha_{j}\bar{\alpha}_{k}\bar{\alpha}_{l}. \label{W}
\end{align}
Here, the coefficients $\Omega_{ijkl}$  are complicated integrals involving AdS mode functions whose form can be recovered by examining Appendix D of \cite{Craps:2014jwa}. The flow equations can be obtained from the effective Lagrangian
\begin{equation}
L=\sum_n i\omega_n\left(\bar\alpha_n\frac{d\alpha_n}{d\tau}-\alpha_n
\frac{d\bar\alpha_n}{d\tau}\right)+2\epsilon^2 W,
\end{equation}
which is invariant under the three continuous symmetries $\alpha_n\to e^{i\omega_n\theta}\alpha_n,\ \alpha_n\to e^{i\theta}\alpha_n,\ \tau\to\tau+\tau_0$, leading to three conserved quantities: the ``free motion energy'' $E=\sum_n\omega_n^2|\alpha_n|^2$ (the conservation of which had already been observed numerically in \cite{Balasubramanian:2014cja}), the ``particle number'' $J=\sum_n\omega_n|\alpha_n|^2$, and the quartic ``interaction energy'' $W$ \cite{Craps:2014jwa} (see also \cite{Basu:2014sia} for the case of a self-interacting probe scalar field). Conservation of $J$ is a direct consequence of $\omega_n=\omega_i+\omega_j-\omega_k$ being the only open flow channels. (Otherwise, (\ref{W}) would contain terms with unequal numbers of $\alpha$s and $\bar\alpha$s, upsetting the symmetry responsible for the conservation of $J$.) We note that these are exact conservation equations of the flow equations, which provide an approximation to the dynamics that is valid on time scales of order $1/\epsilon^2$.

Simultaneous conservation of $E$ and $J$ has immediate consequences \cite{Buchel:2014xwa}. Transferring energy to higher-$n$ modes (which have more energy per particle) tends to decrease $J$. Conservation of $J$ implies that some of the energy must flow to lower-$n$ modes. Such ``dual cascades'' had indeed been observed numerically in \cite{Balasubramanian:2014cja}.

Rewritten in terms of the real amplitudes $a_l$ and phases $b_l$, $\alpha_l\equiv a_le^{-ib_l}/2$, the flow equations read
\begin{equation}
\frac{2\omega_{l}}{\epsilon^{2}}\frac{da_{l}}{dt}
=-\underbrace{\sum_{i}^{\{i,j\}}\sum_{j}^{\neq}\sum_{k}^{\{k,l\}}}_{\omega_{i}+\omega_{j}=\omega_{k}+\omega_{l}}S_{ijkl}a_{i}a_{j}a_{k}\sin(b_{l}+b_{k}-b_{i}-b_{j})
\end{equation}
and
\begin{equation}
\frac{2\omega_{l}a_{l}}{\epsilon^{2}}\frac{db_{l}}{dt}=-T_{l}a_{l}^{3}-\sum_{i}^{i\neq l}R_{il}a_{i}^{2}a_{l}-\underbrace{\sum_{i}^{\{i,j\}}\sum_{j}^{\neq}\sum_{k}^{\{k,l\}}}_{\omega_{i}+\omega_{j}=\omega_{k}+\omega_{l}}S_{ijkl}a_{i}a_{j}a_{k}\cos(b_{l}+b_{k}-b_{i}-b_{j}).
\end{equation}
(The coefficients $S$, $R$, $T$ simply give another parametrization of the $\Omega$ coefficients in (\ref{W}). We generally follow the notation of \cite{Craps:2014vaa, Craps:2014jwa}.) Families of quasiperiodic solutions, which have constant amplitudes, $da_l/dt=0$, have been found numerically by truncating to a finite number of modes \cite{Balasubramanian:2014cja}. These quasiperiodic solutions are thought to be anchors for the islands of stability in the phase space of AdS perturbations \cite{Balasubramanian:2014cja, Buchel:2014xwa}. The abundance of quasiperiodic solutions is intimately related to the missing secular terms in perturbation theory (and therefore to conserved quantities) \cite{Craps:2014vaa, Craps:2014jwa}. The solutions have been explicitly parametrized by conserved quantities and numerically shown to be stable \cite{Buchel:2014xwa, Green:2015dsa}. All this holds for time scales of order $1/\epsilon^2$; it has not been investigated what happens later as that would require going to higher orders of perturbation theory.

Our second main message is as follows: {\em In perturbation theory, the resonant normal mode spectrum of AdS leads to secular terms. Their resummation results in effective flow equations governing the dynamics up to times of order $1/\epsilon^2$ (unless nonlinearities become too strong before that time). Peculiarities of AdS enforce a set of selection rules forbidding certain terms in the effective flow equations. These restrictions generate an extra conservation law and have a number of dynamical consequences.}

An important aspect of our analytic approach is that it brings the short-wavelength regime within reach. We expect that this will be crucial in order to understand the turbulent onset of black hole formation. In fact, as we will briefly describe, it has already led to a fruitful interplay with numerical studies.

First we present a useful tool. The analyticity strip method (introduced in \cite{Sulem}, and in the present context in \cite{Bizon:2013xha}, whose presentation we follow) uses Fourier asymptotics to diagnose singularity formation. Consider the solution $u(t,x)$ of an evolution equation for real-analytic initial data. The analytic extension $u(t,z)$ into the complex plane of the spatial variable will typically have complex singularities moving in time. If a complex singularity hits the real axis, then $u(t,x)$ becomes singular. Denoting the pair of complex singularities closest to the real axis as $z=x\pm i\rho$, so that $\rho$ determines the width of the analyticity strip around the real axis, $u(t,x)$ will be singular if $\rho(t)$ vanishes in finite time. Importantly, $\rho(t)$ is encoded in the $\exp(-\rho k)$ decay of the Fourier coefficients of $u(t,x)$ for large $k$; it can therefore be obtained from the asymptotics of the (numerically determined) Fourier spectrum.

This method has been used to argue that AdS$_3$ is unstable, but not to black hole formation \cite{Bizon:2013xha}. Small perturbations cannot form a black hole because of the energy threshold between AdS$_3$ and the lightest black hole. Numerical evidence suggests that the analyticity strip shrinks exponentially, and does not vanish in finite time; the solution is therefore non-singular (in particular, it has no naked singularity). Nevertheless, AdS$_3$ is unstable because of turbulent flow to short wavelengths. ``Small'' perturbations do not remain ``small.'' Technically, higher Sobolev norms (sums of mode energies in which short-wavelength contributions are preferentially weighted) grow exponentially fast \cite{Bizon:2013xha}, indicating rapid energy transfer to short-wavelength modes.

Extrapolating numerical general relativity to arbitrarily small amplitude $\epsilon$ is involved. In \cite{Bizon:2011gg}, $1/\epsilon^2$ scaling of the collapse time was observed, but does the scaling persist to arbitrarily small $\epsilon$? (See, for instance, \cite{Dimitrakopoulos:2014ada}.) The effective flow equations have $1/\epsilon^2$ scaling built in. If collapse is captured by these equations, then extrapolation becomes possible. A numerical study \cite{Bizon:2015pfa} of the flow equations (truncated to 172 normal modes) for collapsing initial data in AdS$_5$ suggests (for the infinite system) a finite-time oscillatory singularity. A fit to the ansatz
\begin{equation}
a_n(\epsilon^2 t)\sim n^{-\gamma(\epsilon^2 t)} e^{-\rho(\epsilon^2t)n}\ \ (n\gg1),
\end{equation}
motivated by the analyticity strip method, shows indeed that $\rho$ tends to zero in finite time, while $\gamma$ tends to 2, leaving a characteristic power-law spectrum.

Using numerically derived UV asymptotics of the interaction coefficients in the flow equations, it was found in \cite{Bizon:2015pfa} that their amplitude spectrum seemed consistent with the flow equations in AdS$_5$. The UV asymptotics of the interaction coefficients have now been derived analytically for arbitrary dimension \cite{Craps:2015iia}. It remains to be studied how this asymptotics correlates with the power-law spectra emerging from the AdS evolution in general dimensions.

There are a few other recent analytic developments we would like to mention: 
\begin{itemize}
\item Numerical studies of the flow equations suffer from a computation cost bottleneck caused by the complexity of evaluating the interaction coefficients up to high mode numbers. A new recursive method \cite{Craps:2015iia} is hoped to speed this up.
\item Short-wavelength behavior of the interaction coefficients, and in particular its dependence on the number of spatial dimensions, has been studied in \cite{Craps:2015iia}. The results suggest strengthening of turbulent instabilities in higher dimensions.
\item General qualitative features of quasiperiodic solutions have been explained analytically in \cite{Craps:2015xya}. 
\item Spatial dependence of short-wavelength mode functions was considered in \cite{Menon:2015oda} and it was observed that they are strongly peaked near the origin. This peaking furthermore becomes stronger in higher dimensions. These observations are likely to have implications for black hole formation.
\end{itemize}

The current state-of-the-art can be summarized in our third main message: {\em The short-wavelength asymptotics of the time-averaged effective dynamics is particularly relevant for understanding the onset of instability. Analytical and numerical progress have recently been made.}

Given that AdS plays a prominent role in the AdS/CFT correspondence, one might wonder whether its conjectured instability is something to be worried about. The answer is no. The AdS vacuum solution is the lowest energy solution and cannot decay. AdS instability would mean that if one adds an arbitrarily small amount of energy to this vacuum solution, a tiny black hole may form. From a general relativity point of view, a spacetime with a black hole deviates strongly from the vacuum solution, hence the word instability. From a dual field theory perspective, however, instability towards black hole formation simply translates into the statement that arbitrarily small perturbations of the vacuum state can thermalize, which is hardly surprising. From this point of view, gravity solutions that keep oscillating and fail to form a black hole are more intriguing.

Failure of mechanical systems to effectively thermalize in weakly non-linear regimes has been a much-studied subject in non-linear science since the foundational numerical work of Fermi, Pasta and Ulam (FPU), which observed that a system of massive particles connected by weakly non-linear springs does not thermalize efficiently when non-linearities are small. A lack of black hole formation in AdS for certain initial conditions would point to similar phenomena in dual CFTs via the AdS/CFT correspondence. Note that the black hole formation has to be preceded by energy transfer between the bulk AdS modes leading to wave profile focusing. This process itself has similarities to energy transfer driving thermalization in finite-dimensional FPU-like systems \cite{Balasubramanian:2014cja}, though no thermal states exist for classical continuous fields and the process is instead cut off by black hole formation in the bulk (which might be interpreted as genuine thermalization in the dual quantum field theory). Similarly to FPU-like systems, the bulk energy transfer displays some recurrence behavior and the energy flow is not as efficient as one might have hoped, despite the fully resonant linearized spectrum in AdS.

It is often said that small black holes are thermodynamically unstable, in the sense of having lower entropy than a thermal gas at the same temperature, and that they would therefore quickly evaporate if quantum effects were taken into account. It is important to realize, though, that this statement refers to the canonical ensemble, where the system is imagined to be connected to a heat bath. In our setting, we keep the energy fixed, so it is the microcanonical ensemble that is relevant. 
In this ensemble, small black holes will typically evaporate just a tiny bit and reach equilibrium with the resulting radiation \cite{Horowitz:1999uv}. Only in a special parameter regime of tiny black holes, the thermal equilibrium state does not correspond to a small black hole, but rather to a thermal gas; in this case, the formation of a small black hole has been interpreted as pre-thermalization \cite{Dimitrakopoulos:2014ada}. Small black holes in top-down AdS/CFT models do suffer from another type of instability, though. Stringy AdS/CFT backgrounds involve extra bulk dimensions, and small black holes are known to be unstable to localization in the extra dimensions \cite{Hubeny:2002xn}. It has recently been suggested in \cite{Buchel:2015sma} that, while such instability is indeed present for small black holes, it is absent for quasiperiodic solutions.  

One might be worried that in our particular set-up the dynamics was entirely in the scalar field, whereas the metric could be completely integrated out using the constraint equations, by virtue of spherical symmetry. Have we then been able to say anything pertaining to the dynamics of AdS-like geometry, or were all the phenomena we observed mere artifacts related to the specific scalar field dynamics underlying our considerations? It is generally believed that analyzing massless scalar field evolution gives a useful toy model for the full gravitational dynamics, in particular, in relation to the much-studied problems of spherical shell collapse in Minkowski space. There is more direct evidence, however, that the phenomena we have described are by no means due to the specific choice of matter sources that had been deemed convenient. Our need to introduce a scalar field was driven by the absence of any dynamical content in pure spherically-symmetric gravity (one needs to introduce some matter in this setting to obtain any non-trivial dynamics). If the spherical symmetry assumption is relaxed, one can work in pure gravity without any matter sources, and then one is clearly discussing purely gravitational stability of AdS. General non-spherically-symmetric solutions would require dealing with Einstein's equations for functions depending on many variables, which is mostly out of reach of either analytic or numerical methods. Some minimal considerations for this difficult problem have been given in \cite{Dias:2011ss} for a few lowest angular momentum modes at low orders of perturbation theory. The results suggest that turbulent transfer of energy is possible, though in general it is very difficult to proceed further with the analysis. (For non-spherically-symmetric time-periodic solutions, which fail to develop turbulence, a numerical non-perturbative construction has been presented in \cite{Horowitz:2014hja}.) Luckily, in AdS$_5$, for example, one can relax the spherical symmetry in a way that does not require solving Einstein's equations with dependences on more than two coordinates. One can replace the spherical sections in (\ref{metric}) by ``squashed spheres,'' i.e., homogeneous anisotropic spaces of spherical topology (also known as Bianchi IX). In that setting, pure gravity, without scalar fields or any other matter sources, displays non-trivial dynamics, which has been studied numerically and leads in fact to phenomenology nearly identical to the spherically symmetric scalar field case \cite{Bizon}.
 
While one is primarily interested in the full gravitational dynamics on AdS backgrounds, including the pure gravity case we have mentioned, moving in the opposite direction and considering self-interacting probe scalar fields in AdS backgrounds is also interesting from a methodological perspective \cite{Basu:2014sia,BKS2}. Since the AdS background provides a fully resonant spectrum, any non-linearities, and not just gravitational ones, are in a position to have strong effects on the dynamics of arbitrarily small perturbations. In fact, non-linear perturbation theory for the simplest $\phi^4$ scalar field in a non-dynamical AdS background displays strong similarities to the full gravitational case, including the selection rules and conserved quantities \cite{Basu:2014sia,Craps:2014jwa,Yang:2015jha}. The ultimate fate of such probe scalar perturbations seems to be different, however, with the turbulent energy transfer to shorter wavelengths eventually saturating \cite{Bizon2,BKS2}.

Selection rules in AdS perturbation theory \cite{Craps:2014vaa,Craps:2014jwa} are very intriguing, as they must clearly reflect special geometrical properties of the AdS background, but which exact properties those are and how they dictate the selection rules remains unclear. In practice, the selection rules emerge because certain integrals of quartic combinations of AdS mode functions vanish, which is proved directly using the expressions for the mode functions in terms of orthogonal polynomials. For probe scalar fields, the selection rules can be proved with ease, even in the absence of spherical symmetry \cite{Yang:2015jha}. In \cite{Evnin:2015gma}, a hidden $SU(d)$ symmetry was identified, transforming mode functions in AdS$_{d+1}$ into mode functions of the same frequency. The selection rules can be understood in terms of representations of this $SU(d)$ symmetry in which the mode functions featured in the selection rules reside. However, a direct derivation of the selection rules in terms of $SU(d)$ is still lacking, due to algebraic difficulty of constructing the hidden symmetry generators explicitly.

Describing the final stages of collapse remains an outstanding problem and the methods outlined here are not directly applicable. The finite-time singularity in the effective time-averaged description reported in \cite{Bizon:2015pfa} is strongly suggestive of collapse, but the approximations involved in deriving the time-averaged theory (more specifically, the cubic approximation for the effective scalar field non-linearity) strictly speaking break down shortly before the singularity has been reached. It is known that for extremely thin shells, collapse in global AdS is accurately described by the AdS-Vaidya geometry \cite{BM}. It remains to be seen whether and how such description can be incorporated in the full long-time evolution involving gradual focusing and collapse.

Little is known about extending perturbation theory to higher orders and, correspondingly, about the behavior of AdS perturbations of amplitude $\epsilon$ on time-scales longer than $\mathcal{O}(1/\epsilon^2)$. (If the chosen intial data collapse to form a black hole on a time scale of order $1/\epsilon^2$, the question loses its relevance, but not all initial data collapse.) One exception is the time-periodic solutions of \cite{Maliborski:2013jca} which can be constructed iteratively to all orders in perturbation theory. It is believed that these solutions never collapse, but not much is known with certainty about the evolution in their vicinity. The quasiperiodic solutions of \cite{Balasubramanian:2014cja,Green:2015dsa, Craps:2015xya} form a much larger set, including the time-periodic solutions of \cite{Maliborski:2013jca}. These solutions do not collapse on time-scales of order $1/\epsilon^2$, though it is not known what happens afterwards. Further investigations of these different sorts of periodic and quasiperiodic solutions will hopefully elucidate the obstructions to efficient collapse they introduce on different time-scales.

Several groups have studied other implications of AdS (in)stability for thermalization properties of strongly coupled field theories in finite volume:
\begin{itemize}
\item
Oscillations of a shell in AdS result in entanglement entropy oscillations and have been interpreted as revivals of the initial state in field theory \cite{Abajo-Arrastia:2014fma, daSilva:2014zva}. 
\item
Analogies have been drawn to thermalization (or the absence thereof!) in the infinite-volume hard wall model \cite{Craps:2013iaa, Craps:2014eba}. 
\item
The lack of thermalization corresponding to shells that keep oscillating has been suggested to be an artifact of the large $N$ limit \cite{Dias:2012tq}. 
\end{itemize}

Many other developments and open questions are worth mentioning: 
\begin{itemize}
\item
Several of the results we have discussed have been generalized to massive scalar fields coupled to gravity \cite{Kim:2014ida, Okawa:2015xma, Deppe:2015qsa}, to scalar fields in Gauss-Bonnet gravity \cite{Deppe:2014oua} and to a spherical cavity in Minkowski space \cite{Maliborski:2012gx, Maliborski:2014rma, Okawa:2014nea}.  
\item
It is interesting to study the stability of spacetimes with normal mode spectra that are asymptotically resonant \cite{Dias:2012tq, Menon:2015oda}. 
\item
It would be nice to know whether all stability islands are anchored on quasiperiodic solutions \cite{Deppe:2015qsa}. 
\item
Is there a way to prove collapse for arbitrarily small initial data \cite{Bizon:2015pfa}? 
\item
Can one get an analytic handle on the turbulent regime leading to black hole formation? Is there a relation between the power spectra observed in singular blow-ups of solutions to the effective flow equations in \cite{Bizon:2015pfa} and approximate power spectra seen numerically in the strongly turbulent regime in \cite{deOliveira:2012dt}?
\end{itemize}

\noindent\textbf{Acknowledgements:} We would like to thank all collaborators and colleagues who contributed to our work on the subjects presented here,
and especially Joris Vanhoof. B.C.\ thanks A.\ Van Proeyen and his team for a very nice workshop, and the scientific committee for the invitation to speak. The work of B.C.\ has been supported by the Belgian Federal Science Policy Office through the Interuniversity Attraction Pole P7/37, by FWO-Vlaanderen through project G020714N, and by the Vrije Universiteit Brussel through the Strategic Research Program ``High-Energy Physics.'' The work of O.E. is funded under CUniverse research promotion project by Chulalongkorn University (grant reference CUAASC).

\end{document}